\documentclass[prl,amsmath,amsfonts,amssymb,twocolumn,superscriptaddress,showpacs]{revtex4}

\usepackage{graphicx,psfrag}
\usepackage{dcolumn}
\usepackage{bm}
\usepackage{mathbbol}
\usepackage{makeidx}               
\usepackage{graphicx}              
\usepackage[bookmarksnumbered,colorlinks,plainpages,citecolor=blue,urlcolor=blue]{hyperref}
\usepackage[rflt]{floatflt}
\usepackage{subfigure}
\usepackage{caption2}
\usepackage{latexsym}
\usepackage{amsmath}
\usepackage{amsfonts}
\usepackage{amssymb}
\usepackage{enumerate}
\usepackage{epstopdf}

\newcommand{\uck}[1]{\o}

\newcommand{\ket}[1]{\mbox{$|#1\protect\rangle$}}

\newcommand{\bra}[1]{\mbox{$\protect\langle#1|$}}

\def\beq{\begin{equation}}
\def\eeq{\end{equation}}
\def\bea{\begin{eqnarray}}
\def\eea{\end{eqnarray}}

\begin{document}

\begin{titlepage}
\date{\today}
\title{On the Power of Weak Measurements in Separating Quantum States}

\author{Boaz Tamir}
\email{boaz_tamir@post.bezalel.ac.il}
\affiliation{\mbox{Faculty of Interdisciplinary Studies, Bar-Ilan University, Ramat-Gan, Israel}}

\author{Eliahu Cohen}
\email{eliahuco@post.tau.ac.il}
\affiliation{\mbox{School of Physics and Astronomy, Tel Aviv University, Tel Aviv, Israel}}

\author{Avner Priel}
\email{apriel@ualberta.ca}
\affiliation{\mbox{Department of Physics, University of Alberta, Alberta, Canada}}

\maketitle

\section{Abstract}

We investigate the power of weak measurements in the framework of quantum state discrimination. First, we define and analyze the notion of weak consecutive measurements. Our main result is a convergence theorem whereby we demonstrate when and how a set of consecutive weak measurements converges to a strong measurement. Second, we show that for a small set of consecutive weak measurements, long before their convergence, one can separate close states without causing their collapse. We thus demonstrate a tradeoff between the success probability and the bias of the original vector towards collapse. Next we use post-selection within the Two-State-Vector Formalism and present the non-linear expansion of the expectation value of the measurement device's pointer to distinguish between two predetermined close vectors.

\section{Introduction}


Weak measurement \cite{AAV} has already been proven to be very helpful in several experimental tasks \cite{SpinHall,SNR,WhiteLight,Jordan}, as well as in revealing fundamental concepts \cite{RHardy,Inter,Cheshire,Past,Potential}. Tasks traditionally believed to be self contradictory by nature such as determining a particle's state \emph{between} two measurements prove to be perfectly possible with the aid of this technique. Within the framework of the Two-State-Vector Formalism (TSVF), weak measurements reveal several new and sometime puzzling phenomena. For a general discussion on weak measurements see \cite{AAV,Aharonov-Vaidman,Aharonov,Tamir,ACE}. In this paper we analyze the strength of weak measurements by addressing the question of quantum state discrimination.

There are several known methods for quantum state discrimination, i.e. for the task of deciding which vector was chosen out of a predetermined set of (possibly close) state-vectors (for a general review see \cite{Chefles}). Discriminating between predetermined non-orthogonal vectors can be seen as quantum hypothesis testing \cite{Chefles}. Suppose we use projective measurements, then the question is what are the best projections to choose so that the error probability would be as small as possible. This was first discussed by Helstrom in \cite{Helstrom} for the case of two predetermined vectors. The density matrix version of the problem was developed by Osaki in \cite{Osaki}. A different scheme was presented by Ivanovic \cite{Ivanovic} where the discrimination is error free but there is a probability for obtaining a non-conclusive result, i.e. a `don't know' result. A variant of this scheme was suggested in \cite{Bagan}, \cite{Herzog} where the inconclusive outcomes have a fixed rate. Recently, Zilberberg et al. \cite{Zilberberg} have presented a new method, based on partial measurements followed by post selections. The problem of state discrimination and the problem of cloning are deeply connected. A recent paper of Yao et al. \cite{Yao} discusses approximate cloning and probabilistic cloning. Weak measurements were also proposed for the task of discrimination between two very close states \cite{Qiao}. The scheme we suggest here in Ch. II is similar, yet somewhat more general, while the scheme we use at Ch. I, based on an sequential weak measurements is quite different. The advantages and disadvantages of each method will be discussed.

In Ch. I we discuss the orbit of a two dimensional state-vector under the set of transformations induced by weak measurements. This process can be described as a biased Gaussian random walk on the unit circle. The probability amplitude that govern the next step (the `coin' probability amplitude) is changing with each step. In this, our walk is similar to the one presented in \cite{Murch}, but differs from the quantum random walk presented in \cite{Kempe}. We thereby rotate two different vectors in opposite directions (therefore in a non-unitary way), where we can use a single (strong) projective measurement to distinguish between them. We show that using enough weak measurements (the number of which is a function of the weak coupling), the overall success probability converges to the known optimal result for discrimination by projective measurements \cite{Helstrom}.

Next we use a different approach; we reduce the number of measurements, thus compromising the success probability, however, gaining an advantage by avoiding the collapse.

In Ch. II we apply the TSVF of weak measurements. By choosing the right Hermitian operator and a proper post-selection we can get imaginary weak values. Imaginary weak values are best suited for the analysis of the coordinate variable of the measurement space. We apply a non-linear expansion of the weak value and use it to compute the first and second moments of such a variable. These moments of the coordinate variable change as functions of the initial vector. We then pick two state vectors maximizing the difference between the two distributions of the coordinate variable.


\section{Chapter I: A cloning protocol using iterative weak measurements}

In this Chapter we shall perform consecutive weak measurements (without post-selection) to show that it is possible in principle to differentiate between two non-orthogonal vectors. Suppose Alice is sending Bob one of two predetermined state vectors of a two dimensional system $S$:

\[ \ket{\psi_1}= \cos \phi \ket{0} + \sin \phi \ket{1} \]

\[ \ket{\psi_2}= \sin \phi \ket{0} + \cos \phi \ket{1} \]

\noindent where $\ket{0}$  and $\ket{1}$ are the eigenvalues of $S_z$ and $\frac{\pi}{2} >\phi > \frac{\pi}{4}$. Let $\theta$ be the angle between the two vectors:

\[  \bra{\psi_1} \psi_2\rangle = \cos\theta, \]

\noindent therefore the two vectors have the same angle $\frac{\theta}{2}$ with respect to the vector $\frac{1}{\sqrt{2}}( \ket{0} +\ket{1})$ (see Fig. \ref{fig1}). We assume Alice is sending each of the vectors with the same probability. It is well known by \cite{Helstrom} that the maximal success probability is:

\[ P_S(opt) = \frac{1}{2} (1+ \sqrt{1-4\lambda_1 \lambda_2|\bra{\psi_1} \psi_2\rangle |^2})\]

\noindent where Alice is sending $\ket{\psi_1}$ (resp. $\ket{\psi_2}$) with probability $\lambda_1$ (resp. $\lambda_2$). Since we are using $\lambda_1 =\lambda_2 = 1/2$ we can write:

\begin{eqnarray}
P_S(opt) =\frac{1+\sin \theta}{2} = \cos^2 \phi
\end{eqnarray}

Below we shall show that one can reach the same limit using the following weak measurement protocol. By a series of weak measurements Bob will be able to `rotate' the initial vector towards the direction of $\ket{0}$ or $\ket{1}$. Bob will stop the rotations after a predetermined number of iterations, by then he can assume with high probability that the vector would have crossed  $\ket{\tilde{0}}$ or $\ket{\tilde{1}}$ which are close to the axes $\ket{0}$ or $\ket{1}$. He will then strongly measure the final `rotated' vector in the standard $S_z$ basis to get the result $\ket{0}$ or $\ket{1}$. If the result is $\ket{1}$ he can conclude the initial vector was $\ket{\psi_1}$, otherwise it was $\ket{\psi_2}$.\

The error probability has two factors; the first originates from the weak `rotations', i.e. the probability that the weak `rotations' will take $\ket{\psi_1}$ (resp. $\ket{\psi_2}$) to $\ket{\tilde{0}}$ (resp. $\ket{\tilde{1}}$). The second factor originates from the strong final measurement, i.e. the probability that having `rotated' $\ket{\psi_1}$ (resp. $\ket{\psi_2}$) in the correct direction towards $\ket{1}$ (resp. $\ket{0}$) the strong measurements will produce $\ket{0}$ (resp.$\ket{1}$) results.

Below we start by describing the process of weak measurement. Then we describe the protocol in details.

\begin{figure}
\centering
\includegraphics[width=0.99\linewidth]{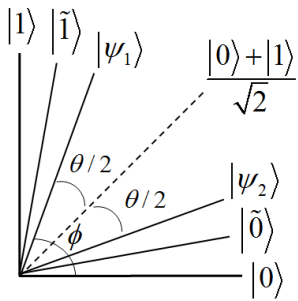}
\caption{The choice of axes, vectors and angles that is used throughout the paper.}
\label{fig1}
\end{figure}


\textbf{I.1. A Gaussian-type random walk induced by weak coupling}




Let $S$ denote our two dimensional system to be measured. Let $\hat{S_z}$ be the Pauli spin matrix on the system $S$. Let $\ket{\psi}= \alpha\ket{0} + \beta\ket{1}$ be a state vector in the eigenbasis $\ket{0}$ and $\ket{1}$ of $\hat{S_z}$.

Let $\ket{\phi}$ denote the wave function of a quantum measurement device. Then,

\begin{eqnarray}
\ket{\phi} = \ket{\phi(x)} = \int_x \phi(x) \ket{x}dx
\end{eqnarray}

\noindent where $\hat{X} \ket{x}= x \ket{x}$ is the position operator of the measuring needle. Suppose $|\phi(x)|^2$ is normally distributed around 0 with variance $ \sigma^2$:

\[ \phi(x) = (2\pi\sigma^2)^{-1/4} e^{-x^2/4\sigma^2}\]

\noindent  The function $\phi(x)$ represents the device's `needle' distribution amplitude. Let $\hat{P}$  be the momentum conjugate operator of the measuring device, such that $[\hat{X},\hat{P}]=i\hbar$.

We shall start the measuring process with the vector:

\[ \ket{\psi} \otimes \ket{\phi(x)} \]

\noindent in the tensor product space of the two systems.  We will now couple the two systems by the interaction Hamiltonian $\hat{H}_{int}$:







\begin{eqnarray}
\hat{H}= \hat{H}_{int}= g(t) \hat{S_z}\otimes \hat{P}
\end{eqnarray}

\noindent where $g(t)$ is the coupling function satisfying:

\[ \int_0^T g(t)dt = g ,\]

\noindent and $T$ is the coupling time. We will use $g=1$ throughout this Chapter for simplicity.

Following the weak coupling the system and the measuring device are entangled:


\begin{eqnarray}
\int_x[\alpha\ket{0}\otimes \phi(x-1)+\beta\ket{1} \otimes \phi(x+1)]\ket{x}dx
\end{eqnarray}

\noindent where the above functions $\phi(x\pm 1)$ are two normal functions with high variance, overlapping each other.


We can write the entangled (unnormalized) state of the measured vector and the measurement device as:

\begin{eqnarray}
\int_{x}  [e^{-\frac{(x-1)^2}{4\sigma^2}}\alpha\ket{0}\otimes \ket{x} + e^{-\frac{(x+1)^2}{4\sigma^2}}\beta\ket{1}\otimes \ket{x}]dx.
\end{eqnarray}

\noindent We will now strongly measure the needle. Suppose the needle collapses to the vector $\ket{x_0}$, then our system is now in the state:

\begin{eqnarray}
\label{bias}
[e^{-\frac{(x_0-1)^2}{4\sigma^2}}\alpha\ket{0} + e^{-\frac{(x_0+1)^2}{4\sigma^2}}\beta\ket{1}] \otimes \ket{x_0}.
\end{eqnarray}

\noindent The eigenvalue $x_0$ could be anywhere around -1 or 1, or even further away, especially if $\sigma$ is big enough i.e. when the measurement is very weak. Note that the collapse of the needle biases the system's vector. However, if $\sigma$ is very large with respect to the difference between the eigenvalues of $\hat{S}_z$ then the bias will be very small and the resulting system's vector will be very similar to the original vector.


Note that the projective measurement on the outer needle's space induces a unitary evolution on the inner space of the composite system. Thus we control the evolution of the state by weak measurements. This resembles an adiabatic evolution of a state vector by strong measurements (see \cite{Childs}).

Consider now the orbit of the initial state vector under the series of weak measurements. We couple the particle to the measuring device and then measure the needle. Next we couple the biased vector to another measuring device and measure its needle. We repeat this process over and over again. The measuring needle is re-calibrated after each measurement, while the particle's state accumulates the successive biases one by one. The series of biased vectors describes an asymmetric random walk on the circle.\





Note that the random walk is continuous and `weighted' in the sense that the distribution function for the next sampling step is changing as a function of the location on the circle. Near the axes $\ket{0}$ and $\ket{1}$ it looks like a Gaussian random walk. The following protocol goes through such a random walk trying to identify the point of start.

\textbf{I.2. Distinguishing by consecutive iterations of weak measurements}

Bob will perform a series of weak measurements to rotate the initial vector $\ket{\psi}$ towards $\ket{0}$ or $\ket{1}$. We use a numerical simulation to investigate the random walk. The vectors $\ket{\tilde{0}}$ and $\ket{\tilde{1}}$ are very close to $\ket{0}$ and $\ket{1}$ respectively, and they will define the `collapse'.

First we address the task of quantifying the number of weak measurements needed to `collapse' the initial vector as a function of the standard error $\sigma$ of the needle.
We simulate the probability distribution of the number of weak measurements needed to collapse the initial vector $\frac{1}{\sqrt{2}} (\ket{0}+\ket{1})$ (see Appendix). We start with $\sigma = 20$. The probability distribution is best fitted ($R^2>0.99$) by a log-normal distribution with $\tilde{\mu} = 2.8$ and $\tilde{\sigma} =0.71$, see Fig. \ref{fig2}. When examining the form of $q_m$ in the Appendix, and applying the Central Limit Theorem under weak dependence, the success of the fit turns obvious: $q_m$ can be described approximately as an exponent of a normally distributed random variable and hence behaves like a log-normal random variable.

\begin{figure}
\centering
\includegraphics[width=0.9\linewidth]{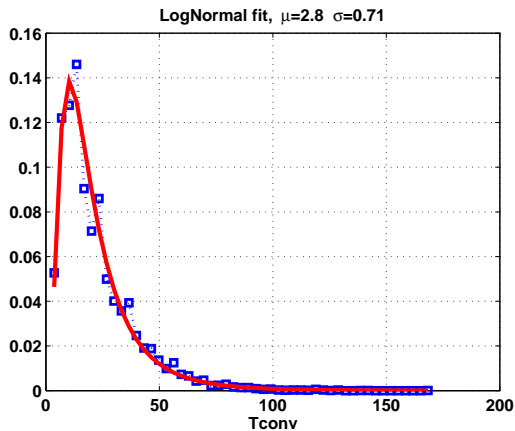}
\caption{Distribution of the number of measurements until the collapse, given a fixed $\sigma$.}
\label{fig2}
\end{figure}

Next we look at the median of the above results (which distribute log-normally) as a function of $\sigma$, see Fig. \ref{fig3}. By changing $\sigma$ we observe a quadratic relation between the median and $\sigma$. In other words, Bob needs $O(\sigma^2)$ steps before he knows the vector had `collapsed' to  $\ket{\tilde{0}}$ or  $\ket{\tilde{1}}$ with high probability. In this simulation we chose $\ket{\tilde{0}}= \cos(10^{\circ}) \ket{0} + \sin(10^{\circ}) \ket{1}$ ; $\ket{\tilde{1}}= \cos(80^{\circ}) \ket{0} + \sin(80^{\circ}) \ket{1}$. The result is independent of the initial vector. Rotating $\ket{\tilde{0}}$ and $\ket{\tilde{1}}$ toward the axes will only multiply the number of steps needed for the collapse by a constant factor.\\

\begin{figure}
\centering
\includegraphics[width=0.9\linewidth]{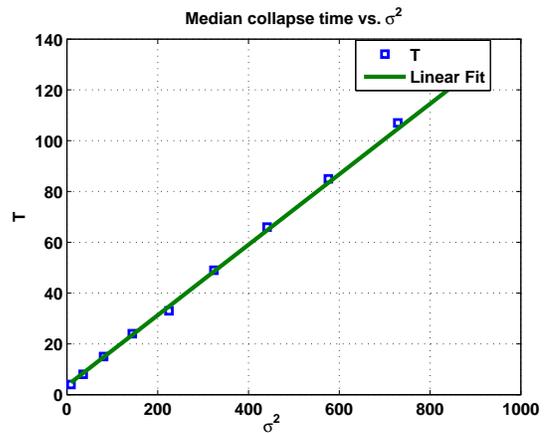}
\caption{Average number of measurements until the collapse as a function of $\sigma$.}
\label{fig3}
\end{figure}

Second we address the error probability in determining the vector's original identity. Having started with $\ket{\psi_1}$, the error probability $Err(\ket{\psi_1})$ is:

\[ Err(\ket{\psi_1})= P^w(\ket{\psi_1}\rightarrow \ket{\tilde{0}}) P^s(\ket{\tilde{0}}\rightarrow \ket{0})+ \]
\[+ P^w(\ket{\psi_1}\rightarrow \ket{\tilde{1}})P^s(\ket{\tilde{1}}\rightarrow \ket{0}).\]

\noindent The error probability for $\ket{\psi_1}$ is the probability that the series of weak measurement iterations will take $\ket{\psi_1}$ to $\ket{\tilde{0}}$ and the strong measurement will take $\ket{\tilde{0}}$ to $\ket{0}$, plus the probability that the weak process will take $\ket{\psi_1}$ correctly to $\ket{\tilde{1}}$, but the strong measurement will take $\ket{\tilde{1}}$ to $\ket{0}$.\

In the next simulation we compute the probability $P^w(\ket{\psi_1}\rightarrow \ket{\tilde{1}})$ and $P^w(\ket{\psi_2}\rightarrow \ket{\tilde{0}})$ for a set of initial vectors $\ket{\psi_i}$ where $\ket{\tilde{0}}=\cos{1}^{\circ} \ket{0} + \sin {1}^{\circ} \ket{1}$ and $\ket{\tilde{1}}=\cos{89}^{\circ} \ket{0} + \sin {89}^{\circ} \ket{1} $. Fig. \ref{fig4} presents the success probability as a function of  $\theta$. The simulation was performed 1000 times (see the pseudo-code in the appendix) where each time we followed the trajectory of the random walk until it crossed the boundaries defined by $\ket{\tilde{0}}$ and $\ket{\tilde{1}}$. The success probabilities are higher than the best separation value in \cite{Helstrom}. However, these simulations disregard the possible error in the strong measurements. As we increase the angle between $\ket{\tilde{0}}$ and $\ket{\tilde{1}}$, we reduce the error of the strong measurement and the success probabilities of the weak process approach the limit in \cite{Helstrom} from above.

\begin{figure}
\centering
\includegraphics[width=0.9\linewidth]{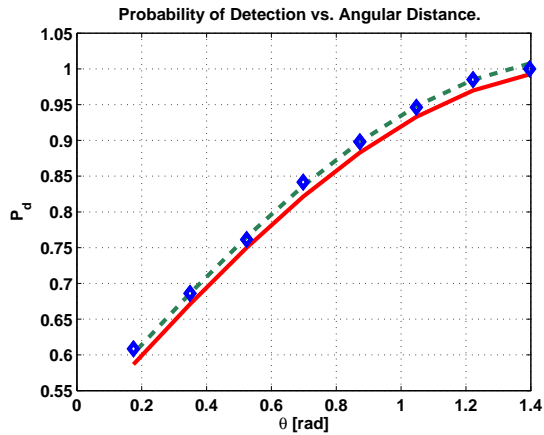}
\caption{The success probability $P^w(\ket{\psi_1}\rightarrow \ket{\tilde{1}})$ as a function of $\theta$. The solid curve describes the optimal success probability for projective measurements, $P^s(\ket{\psi_1}\rightarrow \ket{1})$. Note that the success probabilities are with respect to different outcome vectors.}
\label{fig4}
\end{figure}

The following conclusion is only natural:

\textbf{Conclusion} $Err(\ket{\psi_1})= \sin^2\alpha$.

We can demonstrate the conjecture by extending the angle between $\ket{\tilde{0}}$ and $\ket{\tilde{1}}$.

\textbf{I.3. Hypothesis testing with weak measurements}

So far we have used the weak measurements to iteratively produce small biases of the initial vector to shift it towards one of the axes. The results we got for the pointer of the weak measurement apparatus were so far ignored. Suppose now we use a very small number of weak measurements. We could use the pointers' readings as samples from the vector's distribution. Moreover, we can average over the few values and use the result as a statistic. Notice however that we are sampling from a distribution that is changing following each pointer's reading. The advantage of such a protocol lies in the fact that the vector has not collapsed; if the standard error of the weak measurement is large and the number of weak measurement is small then the resulting vector is still in the neighborhood of the original one. We will show that the distributions of the averages behaves as a function of the initial vector and therefore could be used to distinguish between the two.

In the next simulation (Fig. \ref{fig5}) we weakly measured the vectors for 5, 10, and 20 times, using $\sigma = 3$. This simulation was performed 5000 times for several different angles $\theta$ (as in Fig. \ref{fig1}). It is expected that the average value (of weak measurements) for $\ket{\psi_1}$ (resp. $\ket{\psi_2}$ should be below 0 (resp. above 0). The success probability was computed by the number of times the average value did not cross 0.

\begin{figure}
\centering
\includegraphics[width=0.9\linewidth]{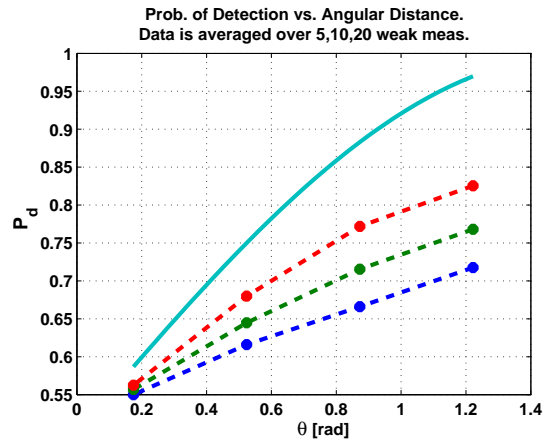}
\caption{Success probability for hypothesis testing with low number of weak measurements. The solid curve describes the optimal success probability for projective measurement (see Eq. 1 above).}
\label{fig5}
\end{figure}

\noindent Fig. \ref{fig6} describes the cumulative distribution function for the above average values (denoted by $x$) for the initial angle $\theta = 50^{\circ}$ and initial vector $\ket{\psi_2}$. It can be seen that the graphs are similar to a shifted cumulative distribution function of a normal random variable. Moreover, the median is fixed regardless of the number of weak measurements. This median is supposed to coincide with the theoretical `strong' one when $\sigma$ becomes lower.

\begin{figure}
\centering
\includegraphics[width=0.9\linewidth]{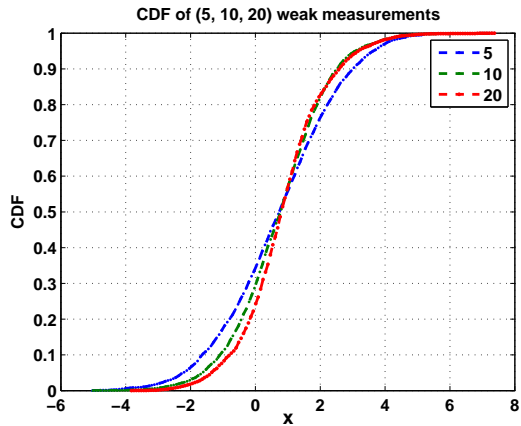}
\caption{Cumulative distribution function for the average of weak values for the initial angle $\theta = 50^{\circ}$ and initial vector $\ket{\psi_2}$.}
\label{fig6}
\end{figure}

In practice, if we use such a protocol to distinguish between two vectors (separated by an angle $\theta$, as in Fig. \ref{fig1}) we should be aware of two types of errors. The first comes from the fact that the vector is changing with each weak measurement. It could possibly drift following the first measurement towards the other vector, thereafter staying in that neighborhood  for long. Thus our readings will identify the wrong vector. The second type of error comes from hypothesis testing. The sample of the average could be very close to 0 making the decision tougher. The simulation above does not distinguish between the two types of errors. It only matches the true vector with the average value of weak measurements, to give an overall success probability.

\section{Chapter II. State discrimination with post-selection: the non-linear computation}

In this Chapter we will use the TSVF of weak measurements to perform a non-linear analysis of quantum state discrimination for a general coupling strength. We will show that for the rare events where the post-selection is successful we have a high probability to identify the vector. The motivation for utilizing post-selection was already verified in many precision measurements \cite{SpinHall,SNR,WhiteLight}, hence it is natural to examine it for this application.

In the pre- and post-selected set of weak measurement (see below) we will look at the distribution of the pointer's coordinate variable $\hat{X}_{fin}$. The moments of $\hat{X}_{fin}$ are functions of the pre-selected state $\ket{\psi_{in}}$, the post- selected state $\ket{\psi_{fin}}$, and the operator $A$. We will show that it is possible to pick two initial vectors $\ket{\psi^i_{in}}$ such that the corresponding distributions of $\hat{X}_{fin}$ are easily distinguished (possibly by one sample). In particular, for one of the initial vectors $\hat{X}_{fin}$ will be distributed around 0 with standard error $\sigma$ while for the other initial vector $\hat{X}_{fin}$ will be distributed around $\sigma$ with very low standard error (almost 0). This will make it easy to differentiate between the two cases.

In II.1 we compute the expansion of the weak value $\bra{\psi_{fin}} e^{-ig \hat{A}\hat{X}/\hbar} \ket{\psi_{in}}$ using all terms in $\hat{A}$ and $\hat{X}$. In II.2 we show how to pick two initial vectors that maximizes the difference between the corresponding distributions of $\hat{X}_{fin}$. Our derivation is based on \cite{Koike}.\\

\textbf{II.1. The non-linear approximation of weak values}

Suppose $\hat{A}$ is an Hermitian operator on the principle system $S$. Let $\ket{\psi}$ denote a state vector for that system. Let $\hbar=1$. We will also assume $\hat{A}^2 =1 $, this will make it easy to write $e^{-ig \hat{A}\hat{X}/\hbar}$ as a power series. Assume also that:

\[ \langle \hat{A} \rangle_w= \frac{\bra{\psi_{fin}}\hat{A} \ket{\psi_{in}}}{\bra{\psi_{fin}}{\psi_{in}}\rangle}= ib\]

Let $\ket{\phi}$ denote the wave function of a quantum measurement device. Then,

\begin{eqnarray}
\ket{\phi} = \ket{\phi(x)} = \int_x \phi(x) \ket{x}dx
\end{eqnarray}

\noindent where $\hat{X} \ket{x}= x \ket{x}$ is the position operator of the measuring needle. We will also assume that $|\phi(x)|^2$ is normally distributed around 0 with variance $\sigma^2$:

\[ \phi(x) = (2\pi\sigma^2)^{-1/4} e^{-x^2/4\sigma^2}\]

\noindent  The function $\phi(x)$ represents the device's `needle' amplitude distribution. We will now couple the principle system and the measurement system by the interaction Hamiltonian $\hat{H}_{int}$:







\begin{eqnarray}
\hat{H}= \hat{H}_{int}= g(t) \hat{A}\otimes \hat{X}
\end{eqnarray}

\noindent (we used the operator $\hat{X}$ instead of $\hat{P}$ since we need imaginary weak values \cite{Jozsa}). Here $g(t)$ is a coupling function satisfying:

\[ \int_0^T g(t)dt = g ,\]

\noindent where $T$ is the coupling time.


We shall start the measuring process with the vector:

\[ \ket{\psi} \otimes \ket{\phi(x)} \]

\noindent in the tensor product of the two systems. Then we apply the Hamiltonian:

\[ e^{-i\hat{A}\hat{X}/\hbar}\ket{\psi} \otimes \ket{\phi(x)} .\]






\noindent Let

\[ \ket{\Phi_{fin}(x)} = \bra{\psi_{fin}} e^{-ig\hat{A}\hat{X}/\hbar} \ket{\psi_{in}} \ket{\phi(x)} \]

\noindent be the wave function of the needle following the coupling and the post-selection. For an observable $\hat{M}$ on the needle's space $\ket{\phi(x)}$, let:

\[ \langle \hat{M}\rangle_{in} = \frac{\bra{\phi} \hat{M} \ket{\phi}}{\bra{\phi} \phi \rangle},\]

\[ \langle \hat{M}\rangle_{fin} = \frac{\bra{\Phi_{fin}} \hat{M} \ket{\Phi_{fin}}}{\bra{\Phi_{fin}} \Phi_{fin} \rangle}.\]

\noindent Since $\hat{A}^2 =1 $ we can write $\bra{\psi_{fin}} e^{-ig \hat{A}\hat{X}} \ket{\psi_{in}}$ as

\[ \sum_{n=0}^\infty \frac{(-ig\hat{X})^{2n}}{(2n)!}\bra{\psi_{fin}}\psi_{in}\rangle + \sum_{n=0}^\infty \frac{(-ig\hat{X})^{2n+1}}{(2n+1)!}\bra{\psi_{fin}}\hat{A} \ket{\psi_{in}}\]

\begin{eqnarray}
= \bra{\psi_{fin}}\psi_{in}\rangle [ \cos(g\hat{X})-i\langle\hat{A}\rangle_w \sin(g\hat{X})]
\end{eqnarray}

\noindent where

\[ \langle \hat{A} \rangle_w= \frac{\bra{\psi_{fin}}\hat{A} \ket{\psi_{in}}}{\bra{\psi_{fin}}{\psi_{in}}\rangle}.\]

\noindent Consider now the average bias of the needle:

\[ \langle\hat{X}\rangle_{fin} = \frac{\bra{\Phi_{fin}}\hat{X} \ket{\Phi_{fin}}}{\bra{\Phi_{fin}} \Phi_{fin}\rangle} \]

\[ = \frac{\langle \hat{X} |\cos(g\hat{X})-i\langle \hat{A}\rangle_w \sin(g\hat{X})|^2\rangle_{in}}{\langle |\cos(g\hat{X})-i \langle \hat{A}\rangle_w \sin(g\hat{X})|^2\rangle_{in}}.\]

\noindent Recall $\langle \hat{A}\rangle_w= ib$, let $a^+= \frac{1+b^2}{2}$ and  $a^- = \frac{1-b^2}{2}$ then

\[ \langle\hat{X}\rangle_{fin}= \]
\begin{eqnarray}
\frac{a^+ \langle\hat{X}\rangle_{in}+ a^-\langle\hat{X}\cos(2g\hat{X})\rangle_{in}+b\langle\hat{X}\sin(2g\hat{X})\rangle_{in}}{a^+ + a^-\langle\cos(2g\hat{X})\rangle_{in}+b\langle\sin(2g\hat{X})\rangle_{in}}
\end{eqnarray}

\noindent \cite{Remark1}. Now since the needle is symmetric (normally distributed) we can write:

\[ \langle \hat{X}\cos(2g\hat{X})\rangle_{in} = \langle\sin(2g\hat{X})\rangle_{in}= \langle \hat{X}\rangle_{in}=0. \]

\noindent Hence

\begin{eqnarray}
\langle \hat{X}\rangle_{fin}= \frac{b \langle \hat{X}\sin(2g\hat{X})\rangle_{in}}{a^++a^-\langle\cos(2g\hat{X})\rangle_{in}}.
\end{eqnarray}

\noindent We shall now use the parametrization $b=\cot(\frac{\eta}{2})$, ($a^-=-\frac{\cos(\eta)}{2\sin^2(\frac{\eta}{2})}$ and $a^+=\frac{1}{2\sin^2(\frac{\eta}{2})}$) therefore:

\begin{eqnarray}
\langle \hat{X}\rangle_{fin}^\eta= \frac{\sin(\eta) \langle \hat{X}\sin(2g\hat{X})\rangle_{in}}{1-\cos(\eta)\langle\cos(2g\hat{X})\rangle_{in}}.
\end{eqnarray}

\noindent Since the needle is Gaussian we can write

\[ \langle\cos(2g\hat{X})\rangle_{in}= e^{-2(g\sigma)^2} \]

\noindent and

\[ \langle \hat{X}\sin(2g\hat{X})\rangle_{in}= 2g \sigma^2e^{-2(g\sigma)^2},\]

\noindent \cite{Remark2} and therefore

\begin{eqnarray}
\langle \hat{X}\rangle_{fin}^\eta= \frac{\sin(\eta)2g \sigma^2}{e^{2(g\sigma)^2}-\cos(\eta)}.
\end{eqnarray}

\noindent The above ratio has maximal value at $\cos(\eta) = e^{-2(g\sigma)^2}$ which is:

\begin{eqnarray}
\langle \hat{X}\rangle_{fin}^{max} = \frac{2g\sigma^2}{\sqrt{e^{4(g\sigma)^2}-1}}.
\end{eqnarray}

\noindent \cite{Remark3} If the measurement is weak, i.e. $g \cdot \sigma \ll 1$, then

\[ \frac{2g\sigma}{\sqrt{e^{4(g\sigma)^2}-1}}\sim 1, \]

\noindent hence

\begin{eqnarray}
\langle \hat{X}\rangle_{fin}^{max} \sim \sigma.
\end{eqnarray}

\noindent This will be true for $\eta$ close to 0. However, if $\eta=0$ then $\langle \hat{X}\rangle_{fin}^\eta =0$ for all $g \neq 0$.

To sum-up $\langle \hat{X}\rangle_{fin}^{\eta}$ is a function of two variables $\eta$ and $g$. If the two variables are correlated such that $\cos(\eta)=e^{-2(g\sigma)^2}$ then for $g$ small enough such that $g\cdot\sigma \ll 1$ we can get $\langle \hat{X}\rangle_{fin}^{\eta}$ to be very close to $\sigma$. But if we fix $g$ (however small) and let $\eta$ go to 0 we can decrease $\langle \hat{X}\rangle_{fin}^{\eta}$ to 0. The variable $\eta$ is a function of the weak values of $\hat{A}$. This means that we can tune $\eta$ such that for $\eta_1$ the expectation value $\langle \hat{X}\rangle_{fin}^{\eta_1}$ will be close to $\sigma$, and for $\eta_2$ the expectation value $\langle \hat{X}\rangle_{fin}^{\eta_2}$ will close to 0.


To compute the variance of the needle after the post-selection note that:\\\\

\begin{eqnarray}
\langle\hat{X}^2\rangle_{fin}=
\end{eqnarray}

\[ \frac{a^+ \langle\hat{X}^2\rangle_{in}+ a^-\langle\hat{X}^2\cos(2g\hat{X})\rangle_{in}+b\langle\hat{X}^2\sin(2g\hat{X})\rangle_{in}}{a^+ + a^-\langle\cos(2g\hat{X})\rangle_{in}+b\langle\sin(2g\hat{X})\rangle_{in}}\]

Since the needle is normally distributed it is easy to see that

\begin{eqnarray}
\langle\hat{X}^2\cos(2g\hat{X})\rangle_{in}= \sigma^2 e^{-2(g\sigma)^2}[1-4g^2\sigma^2],
\end{eqnarray}

\noindent \cite{Remark4} and therefore
\begin{eqnarray}
\langle\hat{X}^2\rangle_{fin}=\sigma^2 \{\frac{a^+ + a^- e^{-2(g\sigma)^2}[1-4g^2\sigma^2]}{a^+ + a^-e^{-2(g\sigma)^2}}\}.
\end{eqnarray}

\noindent For small enough $g\sigma$ the value of $1-4g^2\sigma^2$ will be close to 1 and therefore $\langle\hat{X}^2\rangle_{fin}$ will be close to $\sigma^2$. Note that  $\langle\hat{X}^2\rangle_{fin}$ does not depend on the weak value since the needle has symmetric distribution.

\textbf{II.2. Distinguishing between two non-orthogonal vectors}

We can now pick two small angles $\eta_1$ and $\eta_2$ such that the difference between $\langle\hat{X}\rangle_{fin}^{\eta_1}$ and $\langle\hat{X}\rangle_{fin}^{\eta_2}$ is almost $\sigma$. Since $\eta_1$ and $\eta_2$ correspond to two initial vectors, we can distinguish between the two vectors by estimating the value of $\langle\hat{X}\rangle_{fin}^{\eta}$. In particular, consider

\[ \ket{\psi_{in}^i} = \alpha_i \ket{0} - \beta_i \ket{1} \]

\noindent for i=1,2. Also

\[ \ket{\psi_{fin}} = \frac{1}{\sqrt{2}}(\ket{0} + \ket{1}) \]

\noindent Let $\hat{A}$ be the Hermitian operator:

\[ \hat{A}= \left( \begin{array}{cc} 0&-i\\ i&0 \end{array} \right). \]

\noindent Then $\hat{A}^2=1$. It is easy to see that:

\[ \langle \hat{A} \rangle_w^i = i\frac{\alpha_i+\beta_i}{\alpha_i-\beta_i}\]

\noindent Let:

\[ \eta_1 = arcos( e^{-2(g\sigma)^2}).\]

\noindent Then we can choose $\ket{\psi_{in}^1} = \alpha_1 \ket{0} - \beta_1 \ket{1}$ such that:

\[ \alpha_1 = \frac{1}{\sqrt{2}}(\cos\frac{\eta_1}{2}+\sin\frac{\eta_1}{2})\]

\[ \beta_1 = \frac{1}{\sqrt{2}}(\cos\frac{\eta_1}{2}-\sin\frac{\eta_1}{2})\]

\noindent and therefore

\[ \frac{\alpha_1+\beta_1}{\alpha_1-\beta_1} = \cot(\frac{\eta_1}{2}).\]

\noindent Also, for $\eta_2$ close to 0 we can choose $\ket{\psi_{in}^2} = \alpha_2 \ket{0} - \beta_2 \ket{1}$ such that:

\[ \alpha_2 = \frac{1}{\sqrt{2}}(\cos\frac{\eta_2}{2}+\sin\frac{\eta_2}{2})\]

\[ \beta_2 = \frac{1}{\sqrt{2}}(\cos\frac{\eta_2}{2}-\sin\frac{\eta_2}{2})\]

\noindent hence

\[ \frac{\alpha_2+\beta_2}{\alpha_2-\beta_2} = \cot(\frac{\eta_2}{2}).\]

\noindent Therefore, the difference between $\langle \hat{X} \rangle_{fin}^{\eta_1}$ and
$\langle \hat{X} \rangle_{fin}^{\eta_2}$ will be close to $\sigma$.

\noindent The final variance of the needle in case $\ket{\psi_{in}} =\ket{\psi_{in}^1}$ is

\[ \langle\hat{X}^2\rangle_{fin} - {\langle\hat{X}\rangle_{fin}^{\eta_1}}^2 \approx \sigma^2-\sigma^2 = 0, \]

\noindent and therefore the needle is normally distributed around $\sigma$ with very low standard error.

\noindent The final variance of the needle in case $\ket{\psi_{in}} =\ket{\psi_{in}^2}$ is

\[ \langle\hat{X}^2\rangle_{fin} - {\langle\hat{X}\rangle_{fin}^{\eta_2}}^2 \approx \sigma^2-0 = \sigma^2, \]

\noindent hence the needle is normally distributed around 0 with standard error $\sigma$.

\noindent We can now easily distinguish between the two alternative distributions, possibly with a single sample using standard hypothesis testing (see for example the illustration in Fig. \ref{fig7}).

\begin{figure}
\centering
\includegraphics[width=0.9\linewidth]{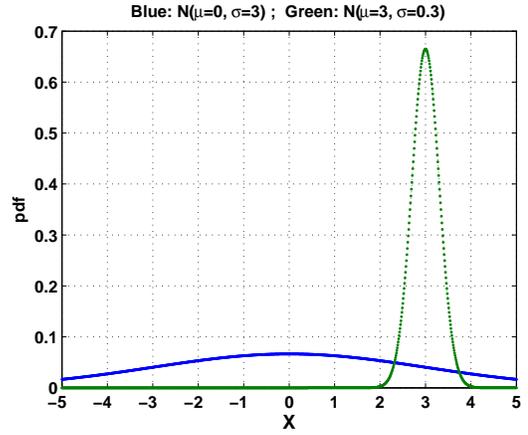}
\caption{Illustration of the normal distribution of the pointer's first moment for each of the two initial vectors.}
\label{fig7}
\end{figure}


The success of the protocol depends solely on the post-selection probability, which is:



\[ |\bra{\psi_{fin}}{\psi_{in}^i}\rangle|^2 = |\frac{\alpha_i-\beta_i}{\sqrt{2}}|^2 = |\sin\frac{\eta_i}{2}|^2. \]

\noindent This probability will be low since each of the initial vectors is almost orthogonal to the final vector.\\

To sum-up, having post-selected the final vector, the probability to correctly guess the right vector could be high. Initially, the probability to post-select is low and therefore the complexity of the protocol depends mainly on the post-selection.

In the above example, the two initial states were represented in the same bases. Including an additional linear transformation in the scheme we can generalized it to the case of two states chosen from two different mutually unbiased bases. This scheme might be suitable for quantum cryptography, and indeed, very recently a method based on sequential weak measurements was suggested for secure key distribution \cite{Troupe}.

\section{Discussion}

Weak measurement theory challenges some of the most basic principles of quantum theory. In a nutshell, it allows the accumulation of information regarding the state-vector without forcing its collapse. As we have shown in Ch. I, when performed many times, weak measurements are equivalent to a single strong one, thus approaching the well-known optimal success probability for discrimination performed by projective measurement. However, when performed only a limited number of times, they do not collapse the vector but only rotate it. In such a case we can still get some weak information about the state by reading (collapsing) the needle of the weak measurement apparatus. Moreover, when followed by post-selection, weak measurements can reveal underlying properties of the initial vector, allowing one to perform quantum state discrimination in retrospect as discussed in Ch. II.\

The gradual process of `getting information while determining the state' which was demonstrated in Ch. I is strictly connected to the old measurement problem \cite{vonNeumann}. By weakly measuring the initial unknown vector we slowly collapse it, creating a continuous tradeoff between our knowledge and its superposition. This process can be thought of as a step-by-step decoherence in which the measured system `leaks' through a small hole (the weak coupling to the measurement device) into the environment. As opposed to traditional de-coherence this process is rigidly controlled and can be stopped at every stage, thus enabling much liberty to the experimenter. Therefore weak measurement's important contribution is its flexibility. With weak measurement one controls the tradeoff between success rate and collapse rate by choosing the strength of the coupling and the number of weak measurement. \

\section{Appendix: A pseudo-code describing the weak orbit }

We wish to find the pointer final position  $q_m$ (the expectation of its distribution) when performing $m$ successive weak measurements on a single particle prepared (for example) in the initial state $\frac{1}{\sqrt{2}}(\ket{0} + \ket{1})$. Let $q_0$ be distributed according to ${N}(1,\sigma^2)$ with probability 1/2 and according to ${N}(-1,\sigma^2)$ with probability 1/2. One can take $\sigma=m \gg 1$. Now let $q_1$ be distributed according to $N(1,\sigma^2)$ with probability:

\small{\[ \frac{exp{[{(q_0 +1)}^2/2\sigma^2]}}{exp{[{(q_0 +1)}^2/2\sigma^2]}+exp{[{(q_0 -1)}^2/2\sigma^2]}}\]}

\normalsize{\noindent and according to $N(-1,\sigma^2)$ with probability:}

\small{\[ \frac{exp{[{(q_0 -1)}^2/2\sigma^2]}}{exp{[{(q_0 +1)}^2/2\sigma^2]}+exp{[(q_0 -1)^2/2\sigma^2]}}.\]}

\normalsize{\noindent Let $q_2$ be distributed according to $N(1,\sigma^2)$ with probability:}

\small{\[ \frac{exp{[((q_0 +1)^2+ (q_1+1)^2)/2\sigma^2]}}{exp[((q_0 +1)^2+(q_1+1)^2)/2\sigma^2]+exp[((q_0 -1)^2+(q_1-1)^2)/2\sigma^2]}\]}

\normalsize{\noindent and according to $N(-1,\sigma^2)$ with probability:}

\small{\[ \frac{exp{[((q_0 -1)^2+ (q_1-1)^2)/2\sigma^2]}}{exp[((q_0 +1)^2+(q_1+1)^2)/2\sigma^2]+exp[((q_0 -1)^2+(q_1-1)^2)/2\sigma^2]}.\]}

\normalsize{Then $q_m$ is distributed according to $N(1,\sigma^2)$ with probability:}

\small{\[ \frac{ exp[ (\sum_{i=0}^{m-1} (q_i+1)^2 )/2\sigma^2]}{exp[ (\sum_{i=0}^{m-1} (q_i+1)^2 )/2\sigma^2]+exp[ (\sum_{i=0}^{m-1} (q_i-1)^2 )/2\sigma^2]}\]}

\normalsize{\noindent and according to $N(-1,\sigma^2)$ with probability:}

\small{\[ \frac{ exp[ (\sum_{i=0}^{m-1} (q_i-1)^2 )/2\sigma^2]}{exp[ (\sum_{i=0}^{m-1} (q_i+1)^2 )/2\sigma^2]+exp[ (\sum_{i=0}^{m-1} (q_i-1)^2 )/2\sigma^2]}.\]}

\normalsize{\section{Acknowledgements}}

We thank Yakir Aharonov, Niv Cohen and Ariel Landau for helpful comments and discussions. E.C was partially supported by Israel Science Foundation Grant No. 1311/14.
\\



\end{titlepage}

\begin{thebibliography}{99}


\bibitem{AAV} Y. Aharonov, D. Albert, L. Vaidman, How the result
of a measurement of a component of the spin of a spin-1/2 particle
can turn out to be 100, Phys. Rev. Lett. 60 (1988) 1351-1354.

\bibitem{SpinHall}
O. Hosten, P. Kwiat Observation of the spin Hall effect of light via weak measurements, Science 319 (2008) 787-790.

\bibitem{SNR}
D.J. Starling,  P.B. Dixon, A.N. Jordan, J.C. Howell, Optimizing the signal-to-noise ratio of a beam-deflection measurement with interferometric weak values, Phys. Rev. A 80 (2009) 041803.

\bibitem{WhiteLight}
X.Y Xu, Y. Kedem, K. Sun, L. Vaidman, C.F. Li, G.C. Guo, Phase estimation with weak measurement using a white light source, Phys. Rev. Lett. 111 (2013) 033604.

\bibitem{Jordan}
A.N Jordan, J. Tollaksen, J.E. Troupe, J. Dressel, Y. Aharonov, Heisenberg scaling with weak measurement: A quantum state discrimination point of view (2014), arXiv:1409.3488.

\bibitem{RHardy}
Y. Aharonov, A. Botero, S. Popescu, B. Reznik, J. Tollaksen, Revisiting Hardy's paradox: counterfactual statements, real measurements, entanglement and weak values, Phys. Lett. A 301 (2002) 130-138.

\bibitem{Inter}
J. Tollaksen, Y. Aharonov, A. Casher, T. Kaufherr, S. Nussinov, Quantum interference experiments, modular variables and weak measurements, New J. Phys. 12 (2010) 013023.

\bibitem{Cheshire}
Y. Aharonov, S. Popescu, D. Rohrlich, P. Skrzypczyk, Quantum Cheshire Cats, New J. Phys. 15 (2013) 113015.

\bibitem{Past}
L. Vaidman, Past of a quantum particle, Phys. Rev. A 87 (2013) 052104.

\bibitem{Potential}
Y. Aharonov,  E. Cohen, S. Ben-Moshe, Unusual Interactions of Pre- and Post-Selected Particles, EPJ Web Conf. {\bf 70} (2014) 00053.

\bibitem{Aharonov-Vaidman} Y. Aharonov, L. Vaidman, The two-state vector
formalism: an updated review. In: Time in Quantum Mechanics, vol.734,
J.G Muga et al. (eds.), Springer Berlin Heidelberg,
(2007) 399-447.

\bibitem{Aharonov} Y. Aharonov, D. Rohrlich, Quantum Paradoxes: Quantum
Theory for the Perplexed, Wiley-VCH, Weinheim (2005).




\bibitem{Tamir} B.Tamir, E.Cohen, Introduction to weak measurements and weak values, Quanta 2 (2013) 7-17.

\bibitem{ACE} Y. Aharonov, E. Cohen, A.C. Elitzur A.C., Foundations and applications of weak quantum measurements, Phys. Rev. A 89,(2014) 052105.

\bibitem{Chefles} A. Chefles, Quantum state discrimination, Contemp. Phys. 41 (2000) 401-424.

\bibitem{Helstrom} C.W. Helstrom, Quantum detection and estimation theory, Academic press, N.Y. (1976).

\bibitem{Osaki} M. Osaki, M. Ban, O. Hirota, Derivation and physical interpretation of the optimum detection operators for coherent-state signals, Phys. Rev. A 54 (1996) 1691.

\bibitem{Ivanovic} I.D. Ivanovic, How to differentiate between non-orthogonal states, Phys. Lett. A 123 (1987) 257-259.

\bibitem{Bagan} E. Bagan, R.Munoz-Tapia, G.A.Olivares-Renteria,J.A.Bergou, Optimal discrimination of quantum states with a fixed rate of inconclusive outcomes, Phys. Rev. A. 86 (2012) 030303.

\bibitem{Herzog} U. Herzog, Optimal state discrimination with a fixed rate of inconclusive results; Analytic solution and relation to state discrimination with a fixed error rate, Phys. Rev. A. 86 (2012) 032314.

\bibitem{Zilberberg} O. Zilberberg, A. Romito, D.J. Starling, G.A. Howland, C.J. Broadbent, J.C. Howell, Y. Gefen, Null values and quantum state discrimination, Phy. Rev. Lett. 110 (2013) 170405.

\bibitem{Yao} G.Chiribella, Y.Yang, A.C.Yao, Quantum replication at the Heisenberg limit, Nat. Commun. 4, (2013) 2915.

\bibitem{Qiao} C. Qiao, S. Wu, Z.B. Chen. Unambiguous discrimination of extremely similar states by a weak measurement (2013), arXiv:1302.5986.

\bibitem{Murch} K.W.Murch, S.J.Weber, C.Macklin, Siddiqi, Observing single quantum trajectories of a superconducting quantum bit, Nature 502 (2013) 211-214.

\bibitem{Kempe} J. Kempe, Quantum random walks: An introductory overview.
Contemp. Phys. 44 (2003) 307-327.

\bibitem{Childs} A. Childs, Quantum information processing in continuous time, Ph.D. Thesis, M.I.T (2004).


\bibitem{Koike} T. Koike, S. Tanaka, Limits on amplification by Aharonov-Albert-Vaidman
weak measurement. Phys. Rev. A 84 (2011) 062106.

\bibitem{Jozsa} R. Jozsa, Complex weak values in quantum measurement,
Phys. Rev. A 76 (2007) 044103.













\bibitem{Remark1}$|\cos(g\hat{X})-i \langle \hat{A}\rangle_w \sin(g\hat{X})|^2= |\cos(g\hat{X})+b \sin(g\hat{X})|^2= \cos^2(g\hat{X})+b^2\sin^2(g\hat{X})+b\sin(2g\hat{X})= (1-b^2)\cos^2(g\hat{X})+ b^2+b\sin(2g\hat{X})=(1-b^2)( \cos(2g\hat{X})+1)/2 +b^2+b\sin(2g\hat{X})=\frac{(1-b^2)}{2}\cos(2g\hat{X})+\frac{(1+b^2)}{2}+b\sin(2g\hat{X})$


\bibitem{Remark2} Use integration by parts on $\int e^{-\frac{x^2}{2\sigma^2}} \cos(kx) dx$

\bibitem{Remark3} Note that $\cos(\eta) = e^{-2(g\sigma)^2}$, and $\sin(\eta)=\sqrt{1-e^{-4(g\sigma)^2}}$, therefore $\frac{\sin(\eta)2g \sigma^2}{e^{2(g\sigma)^2}-\cos(\eta)}= \frac{\sqrt{1-e^{-4(g\sigma)^2}} 2g\sigma^2}{e^{2(g\sigma)^2}(1-e^{-4(g\sigma)^2})}$.

\bibitem{Remark4} Use integration by parts on $\int e^{-\frac{x^2}{2\sigma^2}}[\cos(2kx)-2kx\sin(2kx)]dx$ and  the previous formula for $\langle \hat{X}\sin(2g\hat{X})\rangle_{in}$.

\bibitem{Troupe}
J.E. Troupe, Quantum key distribution using sequential weak values, Quantum Stud.: Math. Found. 1 (2014) 79-96.

\bibitem{vonNeumann} J. von Neumann, Mathematical foundation of quantum mechanics, Princeton University Press (1955).




































































\end{thebibliography}
\end{document}